\documentclass[aps,prb,10pt,superscriptaddress,twocolumn,showpacs]{revtex4-1}
\usepackage{graphicx}
\usepackage{lipsum}
\usepackage{hyperref}
\usepackage{natbib}

\begin{document}

\title{Thermally driven spin and charge currents in thin NiFe$_2$O$_4$/Pt films}

\author{D.\ Meier}
\email{dmeier@physik.uni-bielefeld.de}
\homepage{www.spinelectronics.de}
\author{T.\ Kuschel}
\affiliation{Thin Films and Physics of Nanostructures, Department of Physics, Bielefeld University, D-33501 Germany}

\author{L.\ Shen}
\author{A.\ Gupta}
\affiliation{Center for Materials for Information Technology, University of Alabama, Tuscaloosa Alabama 35487, USA}

\author{T. Kikkawa}
\affiliation{Institute for Materials Research, Tohoku University, Sendai 980-8577, Japan}
 
\author{K. Uchida}
\affiliation{Institute for Materials Research, Tohoku University, Sendai 980-8577, Japan}
\affiliation{PRESTO, Japan Science and Technology Agency, Saitama 332-0012, Japan}
 
\author{E. Saitoh}
\affiliation{Institute for Materials Research, Tohoku University, Sendai 980-8577, Japan}
\affiliation{WPI Advanced Institute for Materials Research, Tohoku University, Sendai 980-8577, Japan}
\affiliation{CREST, Japan Science and Technology Agency, Tokyo 102-0076, Japan}
\affiliation{Advanced Science Research Center, Japan Atomic Energy Agency, Tokai 319-1195, Japan}

\author{J.-M.\ Schmalhorst}
\author{G.\ Reiss}
\affiliation{Thin Films and Physics of Nanostructures, Department of Physics, Bielefeld University, D-33501 Germany}

\date{\today}

\begin{abstract}

We present results on the longitudinal spin Seebeck effect (LSSE) shown by semiconducting ferrimagnetic NiFe\(_2\)O\(_4\)/Pt films from room temperature down to 50\,K base temperature. To the best of our knowledge, this is the first observation of spin caloric effect in NiFe\(_2\)O\(_4\) thin films. The temperature dependence of the conductivity has been studied in parallel to obtain information about the origin of the electric potentials detected at the Pt coverage of the ferrimagnet in order to distinguish the LSSE from the anomalous Nernst effect. Furthermore, the dependence of the LSSE on temperature gradients as well as the influence of an external magnetic field direction is investigated.

\end{abstract}


\maketitle

\section{Introduction}

The spin Seebeck effect (SSE) generates a spin current induced by a temperature gradient and belongs to the emerging field of spin caloritronics \cite{Bauer:2012fq}. The effect was firstly observed in 2008 by Uchida et al. in Ni$_{81}$Fe$_{19}$ (Py) films with Pt stripes \cite{Uchida:2008cc} using an in-plane temperature gradient \(\nabla T\) which generates a spin current perpendicular to the film plane. The spin current was detected in the Pt stripes as an electrommotive force \(\vec{E}_{\text{ISHE}}\) by means of the inverse spin Hall effect (ISHE) \cite{Saitoh:2006kk} given by the formula
\begin{equation}\label{inversespinhalleffect}
	\vec{E}_{\text{ISHE}}=D_{\text{SHE}}\vec{J}_S \times \vec{\sigma}.
\end{equation}
The material-dependent constant \(D_{\text{SHE}}\) describes the magnitude of the spin Hall effect (SHE) \cite{Hirsch:1999wr,Valenzuela:2006tf}, \(\vec{J}_S\) is the spin current and \(\vec{\sigma}\) is the spin polarization vector.

The same effect was also observed in the ferromagnetic semiconductor GaMnAs \cite{Jaworski:2010dy} and in the Heusler compound Co\(_2\)MnSi \cite{Bosu:2011bw}. In 2010 Uchida et al. also demonstrated this effect in the ferrimagnetic insulator LaY\(_2\)Fe\(_5\)O\(_{12}\) (La:YIG) on Gd\(_3\)Ga\(_5\)O\(_{12}\) \cite{Uchida:2010ei} with the same geometric configuration as for the Py/Pt system, which is now referred to as the transversal SSE (TSSE), because the detected spin current is transverse to the applied temperature gradient (\(\vec{J}_S \perp \nabla T\)). Later Uchida et al. presented an alternative configuration for SSE measurements in Y\(_3\)Fe\(_5\)O\(_{12}\) (YIG) \cite{Uchida:2010jb} and Mn-Zn ferrite (Mn,Zn)Fe\(_2\)O\(_4\) \cite{Uchida:2010fb}, the so called longitudinal SSE (LSSE).

For the LSSE, a temperature gradient \(\nabla T\) is applied perpendicular to the film plane in the z-direction (Fig. \ref{lsse_setup_conduct} (a)) and a spin current \(\vec{J}_S\) flows from the magnetic material into a Pt film. The external magnetic field \(\vec{H}\) is aligned in the x-direction, hence the spin polarization vector \(\vec{\sigma}\) also lies in this direction for a sufficiently high  magnetic field. A voltage is measured across the ends of the Pt film along the y-direction (Fig. \ref{lsse_setup_conduct} (a)). Here the detected spin current \(\vec{J}_S\) is aligned parallel to the applied temperature gradient (\(\vec{J}_S \parallel \nabla T\)).

The anomalous Nernst effect (ANE) \cite{Mizuguchi:2012hc} can be observed in the same configuration for ferro- (or ferri-) magnetic metals, which is analog to the anomalous Hall effect (AHE) \cite{Nagaosa:2010js}, but induced by a temperature gradient. The ANE is given by the formula
\begin{equation}\label{nernsteffect}
	\vec{E}_{\text{ANE}}= \alpha \nabla T \times \vec{m}.
\end{equation}
Here, \(\vec{E}_{\text{ANE}}\) is the electromotive force, \(\alpha\) a coefficient describing the magnitude of the ANE, \(\nabla T\) the temperature gradient and \(\vec{m}\) the unit vector of magnetization \cite{Huang:2011cd}. In the LSSE configuration, Eq. (\ref{inversespinhalleffect}) and Eq. (\ref{nernsteffect}) are very similar and in general there has been no effort to separate the longitudinal SSE and the ANE of the measured voltage signal so far \cite{Weiler:2012ti}.

For this reason magnetic insulators like YIG are required to distinguish the voltage signal between longitudinal SSE and ANE, because no ANE is present in insulators. Another promising candidate to study is nickel ferrite (NiFe\(_2\)O\(_4\)). This material grows in the inverse spinel structure consisting of two magnetic sublattices \cite{Li:2011ka}. Recent optical spectroscopy measurements show an indirect bandgap of about 1.6\,eV in the minority channel and a direct bandgap of about 2.4\,-\,2.8\,eV for NiFe\(_2\)O\(_4\) films of thickness between 150 and 270\,nm\cite{Sun:2012jt}. Other experimental data for the optical band gap range from 0.33\,eV to 3.7\,eV for thin films as well as for the bulk\cite{Rai:2012vx,Haetge:2010wk,Dolia:2006tp,Balaji:2005dk,Waldron:1955th}.

Studies relating proximity effects in YIG/Pt structures indicate concern about contributions of the ANE due to spin-polarization of Pt at the interface of the Pt/ferromagnet film \cite{Huang:2012tk,geprags_2012}. Recently, a study of the ANE in YIG/Pt compares the magnitudes of the ANE and LSSE voltage signals by exchanging the temperature gradient \(\nabla T\) and the external magnetic field \(\vec{H}\), but in different configurations\cite{Kikkawa:2012wh}. This study shows that the contribution of the ANE due to proximity effect is less than 5\% of the voltage signal in the LSSE configuration. 

A contribution due to conductivity at the surface of the ferromagnetic film is another possibility to generate an ANE, which can be much greater. This has to be excluded in the first place. In this work the contribution to the measured signal due to a conductive surface of NiFe\(_2\)O\(_4\) films was investigated.

\section{Experimental}

We prepared nickel ferrite (NiFe\(_2\)O\(_4\)) films on 8\,x\,5\,mm\(^2\) MgAl\(_2\)O\(_4\) (100) substrates by direct liquid injection chemical vapor deposition (DLI-CVD) with a thickness of 1.1\,-\,1.2\,\(\mu\)m \cite{Li:2011ka}. In the DLI technique a liquid solution of two precursors (Ni(acac)\(_2\) and Fe(acac)\(_3\)) in the molar ratio of 1:2 and the solvent N, N-dimethyl formamide (DMF) are vaporized at a temperature of 175\(^\circ\)C to obtain a film growth rate in a range of 10\,-\,18\,nm/min at a deposition temperature of 600\(^\circ\)C. In another step a 10nm Pt film was deposited by DC magnetron sputtering in a vacuum system with base pressure of \(5\cdot 10^{-6}\)\,mbar. In an Ar atmosphere with a pressure of \(5\cdot 10^{-2}\)\,mbar a deposition rate of 4.2\,nm/min was obtained.

\begin{figure}[t!]%
\includegraphics[width=\linewidth]{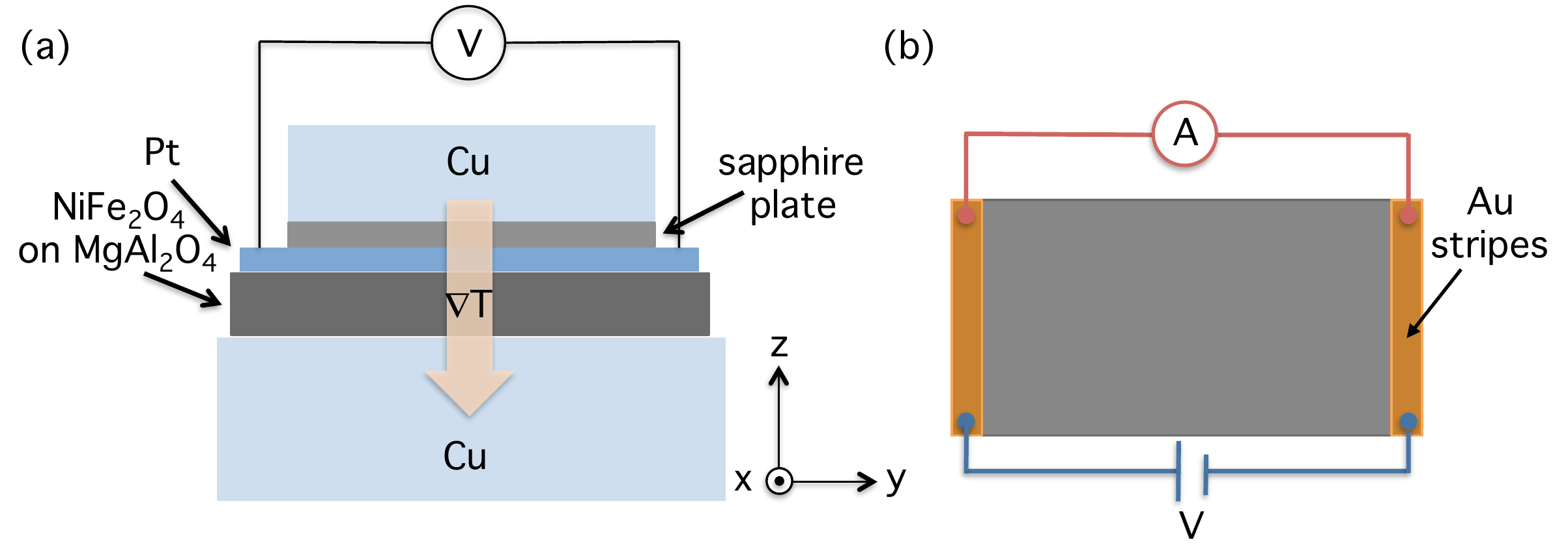}%
\caption{(a) LSSE setup. The sample is sandwiched between two copper plates and a temperature gradient \(\nabla T\) is applied. (b) The geometry for 4-point-measurement. Two sputtered Au stripes establish electric contact to the ends of the sample to hold two lines on the same electrical potential.}%
\label{lsse_setup_conduct}%
\end{figure}%

For the temperature gradient dependent measurements the NiFe\(_2\)O\(_4\)/Pt sample was clamped between two copper plates \cite{Uchida:2012vb}. The lower copper plate was connected to a Peltier thermoelectric module to heat or cool the sample from below, which is electrically insulated from the copper by the substrate. The upper copper plate is connected to a heat bath leading to a heat flow through the sample to generate a temperature gradient \(\nabla T\) in the z-direction (Fig. \ref{lsse_setup_conduct} (a)). The sample is electrically insulated from the upper copper plate by a thin sapphire sheet. The temperature difference was measured between the upper and lower copper plates with two connected T-type thermocouples and were stabilized at  a certain \(\Delta T = \left|(0,0,\Delta T)\right| \propto \nabla T\). Two tungsten needles were contacted at the ends of the Pt film with a micro-probing system along the y-direction. The voltage signal between the two needles was measured with a Keithley 2182A Nanovoltmeter. An external magnetic field \(\vec{H}\) was applied in the x-direction (here defined as \(\alpha = 90^\circ\)) and varied in the range of \(\pm1000\)\,Oe. Afterwards the angle \(\alpha\) between \(\vec{H}\) an the y-direction was varied from \(\alpha = 0^\circ\) to \(\alpha = 360^\circ\) for angle dependent measurements.

In a similar setup the sample was clamped between two copper blocks in a cryostat for low temperature measurements. It was possible to heat both sides separately to generate a temperature gradient perpendicular to the sample plane. Two copper wires were bonded with indium pads on top of the Pt film and connected with a Nanovoltmeter. In the same way a pure NiFe\(_2\)O\(_4\) film without Pt was connected for reference measurements. The lower temperature was measured by a 27\,\(\Omega\) rhodium iron resistance thermometer for temperatures down to 1.5\,K. The temperature on the upper side was measured by a thermocouple. The voltage signal of the thermocouple was calibrated with the rhodium iron sensor. In this setup the copper blocks are also electrically isolated by the substrate and by a sapphire sheet.

Furthermore, the conductivity of the NiFe\(_2\)O\(_4\) film without Pt was characterized in a 4-point-geometry measurement for different temperatures. Fig. \ref{lsse_setup_conduct} (b) shows a sample with two Au-stripes at the ends of the film to hold two point contacts at the same electrical potential. The resistance was measured with a Keithley 2000 in the 4-wire resistance measurement mode. For higher resistance values a voltage of 20\,V was applied and the current was detected with a Keithley 6487 Picoammeter in a 2-point-geometry.

\section{Results}

\begin{figure}[t!]%
\includegraphics[width=\linewidth]{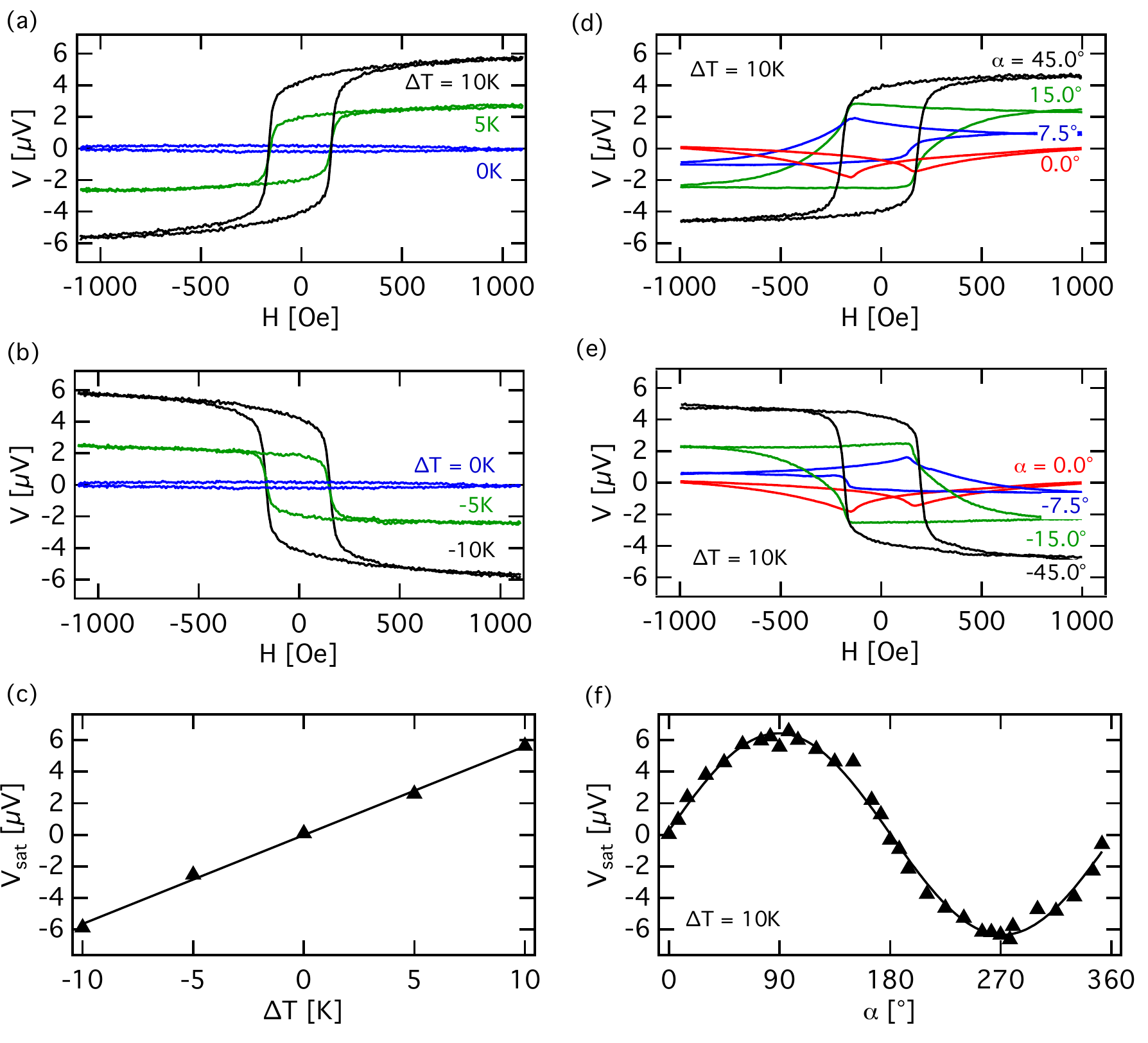}%
\caption{(a), (b) voltage \(V\) as a function of the external magnetic field \(H\) for various temperature differences \(\Delta T\) and \(\alpha = 90^\circ\) at room temperature. (c) the magnitude \(V_{\text{sat}}\) is proportional to the temperature difference \(\Delta T\). (d), (e) voltage \(V\) as a function of \(H\) for various angles \(\alpha\) between \(\vec{H}\) and the y-direction for \(\Delta T = 10\,\text{K}\) at room temperature. (f) the magnitude \(V_{\text{sat}}\) depending on \(\alpha\) for \(\Delta T = 10\,\text{K}\).}%
\label{AG050312b_abcdef2}%
\end{figure}%

\begin{figure}[t!]%
\includegraphics[width=\linewidth]{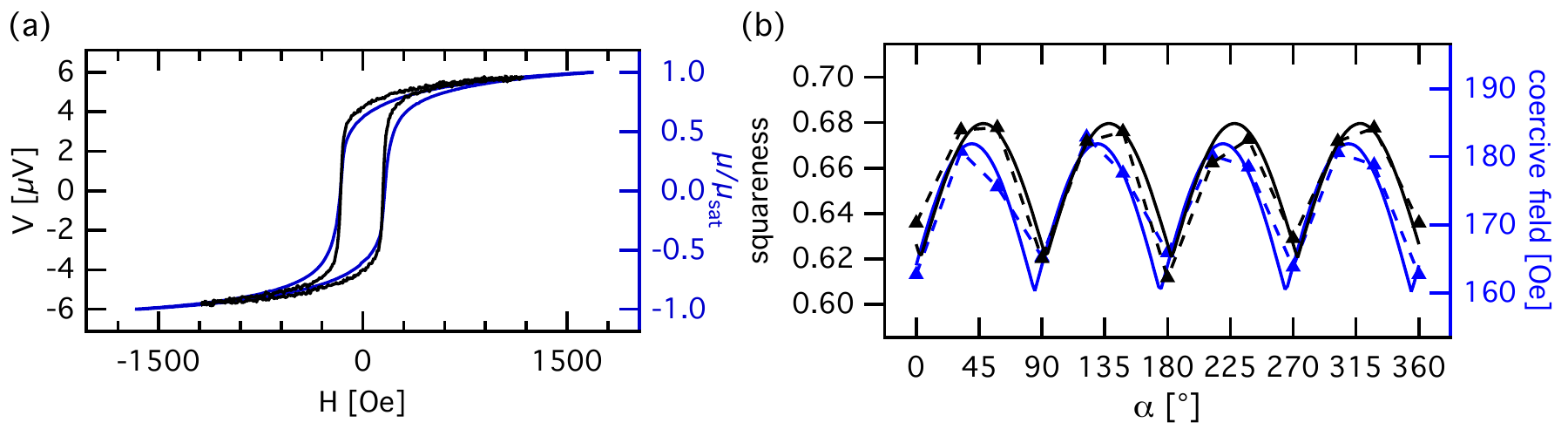}%
\caption{(a) voltage \(V\) as a function of the external magnetic field \(H\) for a temperature difference \(\Delta T=10\text{K}\) (left axis) and the magnetic moment \(\mu\) measured via VSM normalized to the saturation value \(\mu_{\text{sat}}\) (right axis) at room temperature. (b) squareness and coercive field as a function of the angle \(\alpha\) between the external magnetic field \(\vec{H}\) and the y-direction at room temperature. The solid and the dashed curves are lines to guide the eye.}%
\label{AG050312b_moment_remanence_coercivity}%
\end{figure}%

\begin{figure}[t!]%
\includegraphics[width=\linewidth]{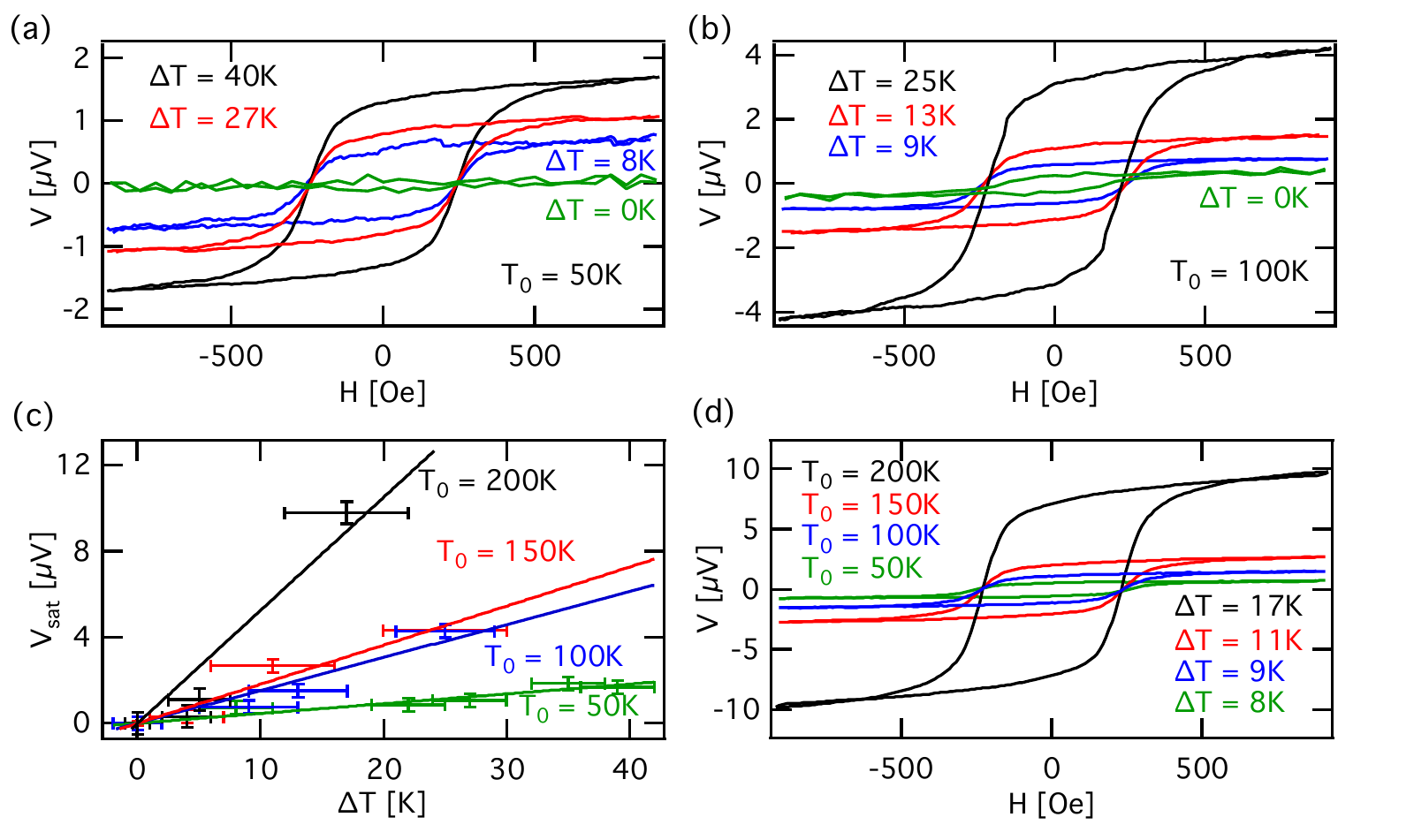}%
\caption{(a),(b) voltage \(V\) as a function of the external magnetic field \(H\) for the mean temperature \(T_0\,=\,50\,\text{K}\) and \(T_0\,=\,100\,\text{K}\) for various temperature differences \(\Delta T\). (c) voltage \(V\) in saturation (\(V_{\text{sat}}\)) as a function of \(\Delta T\) for different mean temperatures \(T_0\). The slope of the linear fits estimate the ratio \(V_{\text{sat}}/\Delta T\) for a certain mean temperature \(T_0\). (d) voltage \(V\) as a function of \(H\) for different mean temperatures \(T_0\) and \(\Delta T\,=\,(12\,\pm\,5)\,\text{K}\).}%
\label{tiefe_Temperaturen}%
\end{figure}%

\begin{figure}[t!]%
\includegraphics[width=\linewidth]{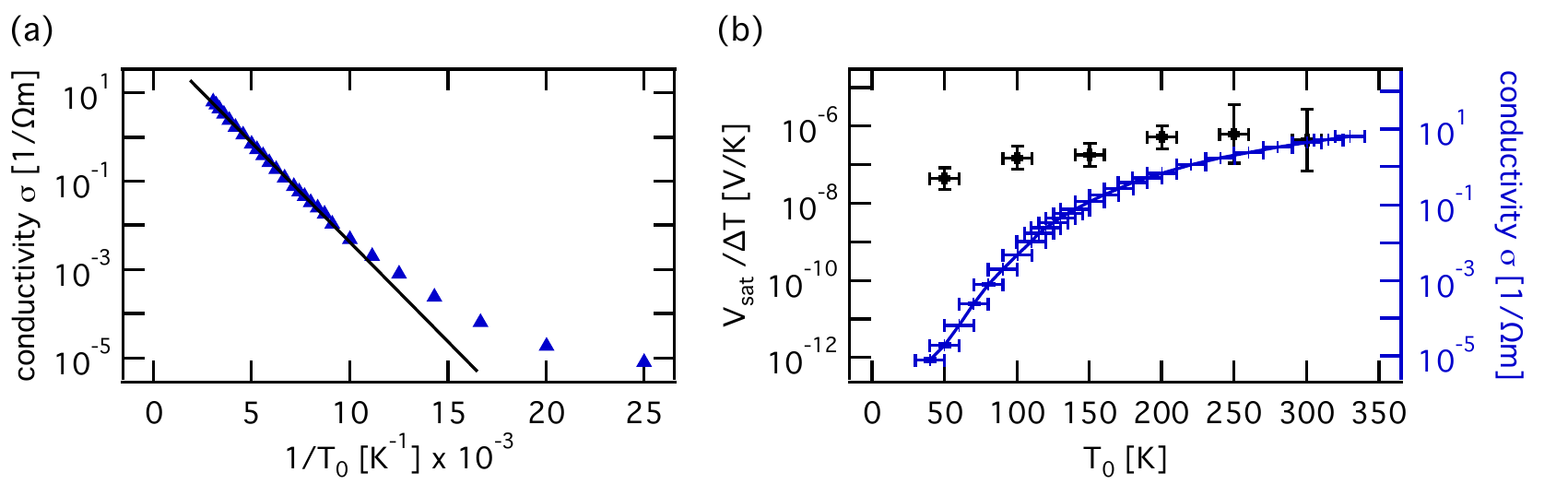}%
\caption{(a) conductivity \(\sigma\) of a NiFe\(_2\)O\(_4\) film as a function of the inverted temperature \(1/T_0\) which shows a semi-conductive behavior and the linear fit in the intrinsic region with a slope of about (-1031\,\(\pm\)\,8)\,K\(\cdot\Omega^{-1}\text{m}^{-1}\). (b) the conductivity \(\sigma\) and the ratio between the saturated voltage signal \(V_{\text{sat}}\) and the temperature difference \(\Delta T\) as a function of the mean temperature \(T_0\).}%
\label{conductivity_Vsat}
\end{figure}%

\begin{figure}[t!]%
\includegraphics[width=\linewidth]{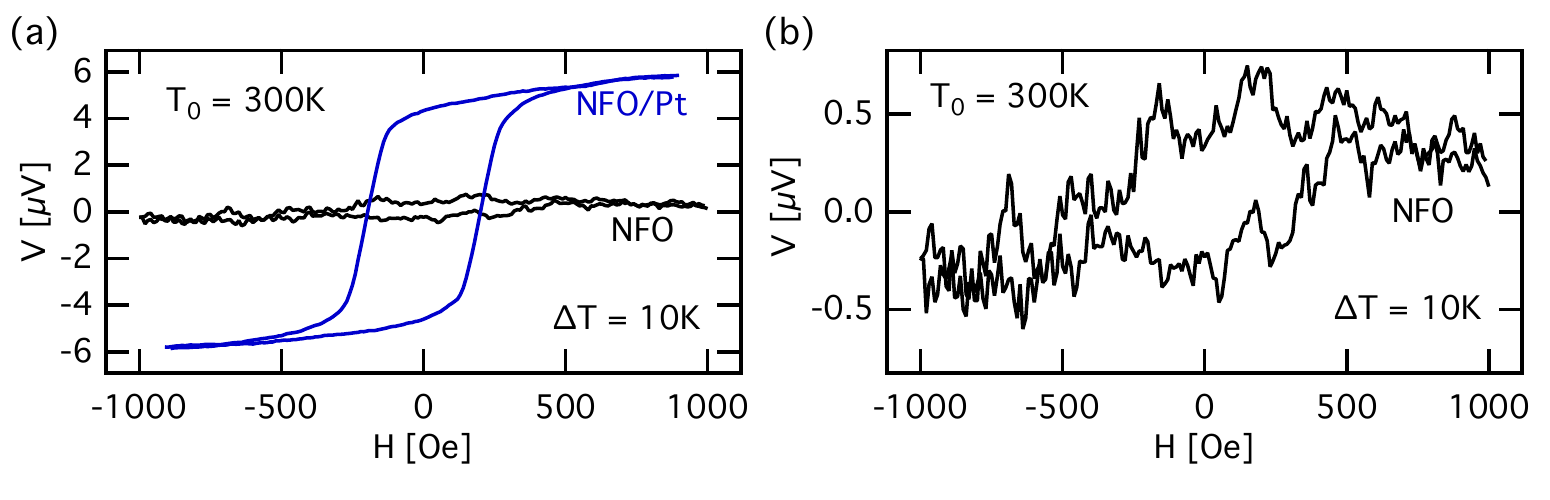}%
\caption{(a) voltage \(V\) as a function of the external magnetic field \(H\) for \(T_0 = 300 \text{K}\) and \(\Delta T = 10 \text{K}\) for a pure NiFe\(_2\)O\(_4\) film (NFO) and a  NiFe\(_2\)O\(_4\)/Pt system (NFO/Pt). (b) the curve for the NFO film in a larger scale.}%
\label{NFO_ANE_NFOPt_300K_delta10K}
\end{figure}%

Fig. \ref{AG050312b_abcdef2} (a) and Fig. \ref{AG050312b_abcdef2} (b) show the voltage signal \(V\) at the ends of the Pt film as a function of the external magnetic field \(\vec{H}\). An asymmetric behavior is observed which can be compared to measurements from other groups on YIG/Pt films\cite{Uchida:2010jb,Weiler:2012ti} which had the same angle \(\alpha=90^\circ\) between \(\vec{H}\) and the y-direction. The magnitude \(V_{\text{sat}}\) of these curves (saturation value) defined by \((V_{\text{max}}-V_{\text{min}})/2\) is proportional to the temperature difference \(\Delta T\) (Fig. \ref{AG050312b_abcdef2} (c)). 

The voltage signal \(V_{\text{sat}}\) also depends on the angle \(\alpha\). Fig. \ref{AG050312b_abcdef2} (f) shows a sinusoidal shape where \(V_{\text{sat}}\) vanishes at \(\alpha = 0^\circ\) and 180\(^\circ\) and has its maximum value at \(\alpha = 90^\circ\) and \(\alpha = 270^\circ\). Here, we emphasize, that the line drawn in Fig. \ref{AG050312b_abcdef2} (f) is a pure sine-function. Thus, the angle dependence of \(V\) perfectly corresponds to the cross-product of Eq. (\ref{inversespinhalleffect}) and Eq. (\ref{nernsteffect}). In Fig. \ref{AG050312b_abcdef2} (d) and \ref{AG050312b_abcdef2} (e) measurements of \(V\) as a function of \(\vec{H}\) with a fixed angle \(\alpha\) are presented. The magnitude of \(V_{\text{sat}}\) gets smaller with decreasing angle \(\alpha\) and the asymmetric behavior nearly vanishes for \(\alpha = 0^\circ\). An explanation for the symmetric contribution for \(\alpha = 0^\circ\) could be the magnetic anisotropy of the NiFe\(_2\)O\(_4\) films as was also demonstrated for YIG films \cite{Weiler:2012ti}. 

Therefore, measurements using vibrating sample magnetometry (VSM) are performed and presented in Fig.\,\ref{AG050312b_moment_remanence_coercivity}. The observed magnetic moment \(\mu\) normalized to the saturation value \(\mu_{\text{sat}}\) is compared in Fig.\,\ref{AG050312b_moment_remanence_coercivity}\,(a) to the obtained voltage signal \(V\) for \(\Delta T = 10\text{K}\) at room temperature. One can see that the coercive field of about (165\,\(\pm\)\,25)\,Oe is identical with the coercive field of the voltage signal. In Fig. \ref{AG050312b_moment_remanence_coercivity} (b) the squareness (\(\mu_{\text{rem}}/\mu_{\text{sat}}\)) and the coercive field of a pure NiFe\(_2\)O\(_4\) film is shown as a function of the angle \(\alpha\). Two magnetic easy axes are obtained induced by the fourfold magnetic anisotropy of the cubic inverse spinel structure of the NiFe\(_2\)O\(_4\) films. The magnetic easy axes are aligned in \(\alpha=45^\circ/225^\circ\) and \(\alpha=135^\circ/315^\circ\) directions due to maximum squareness and coercive field. These directions correspond to the \(<110>\) directions of the crystallographic NiFe\(_2\)O\(_4\) structure. Hence, for \(\alpha =0^\circ\) the voltage signal for small external magnetic fields \(\vec{H}\) gets a symmetric contribution as already discussed for Fig. \ref{AG050312b_abcdef2} (d) and \ref{AG050312b_abcdef2} (e)  when the magnetic moment of the NiFe\(_2\)O\(_4\) changes its in-plane orientation and flips into the direction of the next magnetic easy axis. For a complete interpretation of the symmetric contribution further investigations are required.

Figs. \ref{tiefe_Temperaturen} (a) and (b) show the voltage \(V\) as a function of the external magnetic field \(\vec{H}\) for the mean temperatures \(T_0\,=\,50\,\text{K}\) and \(T_0\,=\,100\,\text{K}\) and for various temperature differences \(\Delta T\). The voltage signal in saturation \(V_{\text{sat}}\) increases for higher temperature differences \(\Delta T\). This is comparable to Fig. \ref{AG050312b_abcdef2} (a) and Fig. \ref{AG050312b_abcdef2} (b) at room temperature, but the proportionality between \(V_{\text{sat}}\) and \(\Delta T\) is less obvious because of the large experimental error in the determination of \(\Delta T\) at low mean temperatures in this setup. In Fig. \ref{tiefe_Temperaturen} (c) \(V_{\text{sat}}\) is shown as a function of \(\Delta T\) for various mean temperatures \(T_0\). For each \(T_0\) the slope of the fit curve describes the ratio \(V_{\text{sat}}/\Delta T\). For the fit curve a linear function (\(y=a\cdot x\)) was assumed considering that for \(\Delta T\,=\,0\,\text{K}\) the voltage signal should vanish. The obtained slopes decrease for lower mean temperatures. In Fig. \ref{tiefe_Temperaturen} (d) the voltage \(V\) is shown as a function of the external magnetic field \(\vec{H}\) for various mean temperatures \(T_0\). The temperature difference is in the range of \(\Delta T\,=\,8\,\text{K}\) and \(\Delta T\,=\,17\,\text{K}\) and an experimental error of  \(\pm\, \Delta T/2\) has to be assumed.

The conductivity \(\sigma\) of a pure NiFe\(_2\)O\(_4\) film (Fig. \ref{conductivity_Vsat} (a)) shows a linear slope with \(1/T_0\) \cite{jefferson:1968cc,vanuitert:1955:lg,vanuitert:1956:lg} which corresponds in a semiconductor model with a band gap of (0.18\,\(\pm\)\,0.01)\,eV and is smaller than the expected value of 1.6\,-2.4\,eV \cite{Sun:2012jt}. Perhaps, the smaller band gap originates from impurities in the sample. The estimation of the band gap from conductivity measurements is sensitive to free charge carriers, which can cause the ANE. Therefore, this technique is suitable to estimate the strength of the ANE. The conductivity decreases from \(\sigma\,=\,4.54\,\Omega^{-1} \text{m}^{-1}\) for \(T_0\,=\,300\,\text{K}\) to \(\sigma\,=\,19.2\times10^{-6}\,\Omega^{-1} \text{m}^{-1}\) for \(T_0\,=\,50\,\text{K}\). This suggests that for high temperatures an electrical conduction in the NiFe\(_2\)O\(_4\) films and therefore possible occurrence of ANE is present. 

Fig. \ref{conductivity_Vsat} (b) compares the conductivity~\(\sigma\) as a function of the temperature \(T_0\) with the ratio of the saturation voltage signal \(V_{\text{sat}}\) and the temperature difference \(\Delta T\) also as a function of \(T_0\). The curves strongly differ for low temperature \(T_0\) with the conductivity decreasing much faster than \(V_{\text{sat}}/\Delta T\). Therefore, the ANE due to the conductivity of the NiFe\(_2\)O\(_4\) film can be neglected at low temperatures. Even by assuming an error in the temperature difference of \(\pm\, \Delta T/2\) and \(\pm 10\,\text{K}\) for \(T_0\), the measured values of \(V_{\text{sat}}/\Delta T\) cannot be explained by the decrease of the conductivity \(\sigma\) with \(T_0\).

In Fig. \ref{NFO_ANE_NFOPt_300K_delta10K} (a) and \ref{NFO_ANE_NFOPt_300K_delta10K} (b) the voltage signal for a pure NiFe\(_2\)O\(_4\) film at room temperature is presented. A small hysteresis is obtained which corresponds to the ANE in the absence of a Pt film on top (Fig. \ref{NFO_ANE_NFOPt_300K_delta10K} (b)), but the magnitude of the voltage signal in saturation is about 10 times smaller than the signal for the NiFe\(_2\)O\(_4\)/Pt system. The small signal to noise ratio is due to the strong temperature dependence of the conductivity of the NiFe\(_2\)O\(_4\) film. Small variations in the temperature difference cause large variations in the offset voltage compared to the voltage signal due to the investigated effects. Detailed measurements at low temperature are required. However, for low temperatures the conductivity of the pure NiFe\(_2\)O\(_4\) decreases very rapidly. 

For a quantitative comparison of the ratio of the saturation voltage signal \(V_{\text{sat}}\) and the temperature difference \(\Delta T\), Kikkawa et al. introduce the ratio \(\tilde{V}_{\text{sat}}/\Delta T\) including geometric factors of the sample\cite{Kikkawa:2012wh}. This coefficient is given by the formula \(\tilde{V}_{\text{sat}} = V_{\text{sat}} L_z/L_y\), where \(L_z\) is the sample thickness in the direction of the temperature gradient and \(L_y\) is the length of the sample in the y-direction. The sample thickness is dominated by the substrate and is about \(L_z\)\,=\,500\,\(\mu\)m. The distance between the voltage contacts is about \(L_y\)\,=\,8\,mm. For \(T_0 = 300\text{K}\) a value of \(\tilde{V}_{\text{sat}}/\Delta T = (30\pm20)\,\text{nV/K}\) is found which is an order of magnitude smaller than the values Kikkawa et al. found for YIG/Pt with \(\tilde{V}_{\text{sat}}/\Delta T = 220\,\text{nV/K}\). \cite{Kikkawa:2012wh} This coefficient decreases to \(\tilde{V}_{\text{sat}}/\Delta T = (2.8\pm0.7)\,\text{nV/K}\) for \(T_0 = 50\text{K}\) in the NiFe\(_2\)O\(_4\)/Pt system. Uchida et al. have shown that for single-crystalline YIG/Pt an enhanced magnitude of  \(V_{\text{sat}}/\Delta T\) around \(T_0\,=\,50\,\text{K}\) appears but not for polycrystalline YIG/Pt.\cite{Uchida:2012vb} An enhanced magnitude of \(V_{\text{sat}}/\Delta T\) could not be observed for NiFe\(_2\)O\(_4\)/Pt using temperature steps of 50\,K for \(T_0\).

Fig. \ref{NFO_ANE_NFOPt_300K_delta10K} (a) shows that the contribution of the ANE in the NiFe\(_2\)O\(_4\) film is very small even when the NiFe\(_2\)O\(_4\) film exhibits a finite electrical conduction. Therefore, the ANE in the NiFe\(_2\)O\(_4\) is not the origin of the strong temperature dependence which is shown in Fig. \ref{conductivity_Vsat} (b). One possible explanation are proximity effects at the NiFe\(_2\)O\(_4\)/Pt interface as described by Kikkawa et al. for YIG/Pt\cite{Kikkawa:2012wh}. Another possible origin is a contribution from the spin-dependent Seebeck effect\cite{Bauer:2012fq}. Spin-polarized conduction electrons can also generate a spin current which is detected due to the ISHE in the Pt film. This effect disappears in the insulating region of the NiFe\(_2\)O\(_4\) film at low temperatures.

\section{Conclusion}

In conclusion, we investigated NiFe\(_2\)O\(_4\)/Pt films in the LSSE/ANE configuration and found similar results as previously observed for YIG/Pt films \cite{Uchida:2010jb}. A systematic study of the origin of the effect was carried out by varying the temperature gradient and the angle \(\alpha\) between the external magnetic field and the voltage contacts. We observed that the measured voltage signal varies proportional to the applied temperature gradient and is sinusoidal with the angle \(\alpha\). Conductivity measurement of a pure NiFe\(_2\)O\(_4\) film shows semiconducting behavior. Therefore, the observed voltage in the LSSE configuration is explained as a superposition of two effects. On the one hand, the LSSE caused by a generated spin current and detected by the ISHE in the Pt. On the other hand, the ANE caused by the conductive behavior of the magnetic NiFe\(_2\)O\(_4\) film, which should increase with temperature similar to the conductivity. Therefore, conductivity measurements at low temperatures were carried out. A voltage signal and a hysteretical behavior was observed at low temperatures as well as at room temperature, but with a smaller magnitude. A comparison between the ratio of the saturation voltage signal and the temperature difference on the one hand and the temperature dependence of the conductivity of the NiFe\(_2\)O\(_4\) on the other hand shows a divergence between the two curves. Therefore, contribution of the ANE due to the conductivity of the NiFe\(_2\)O\(_4\) at low temperature can be neglected.

Furthermore, measurements of the ANE on NiFe\(_2\)O\(_4\) films without Pt showed a voltage signal 10 times smaller than the signal obtained with Pt. Thus, we conclude that the voltage induced in the Pt is mainly due to a spin current from the NiFe\(_2\)O\(_4\) into the Pt film.

\section{Acknowledgements}

This work was supported by PRESTO-JST ``Phase Interfaces for Highly Efficient Energy Utilization'', CREST-JST ``Creation of Nanosystems with Novel Functions through Process Integration'', a Grant-in-Aid for Research Activity Start-up (24860003) from MEXT, Japan, a Grant-in-Aid for Scientific Research (A) (24244051) from MEXT, Japan, LC-IMR of Tohoku University, the Murata Science Foundation, and the Mazda Foundation. The work at the University of Alabama was supported by NSF-ECCS Grant No. 1102263. Furthermore, the authors gratefully acknowledge financial support by the Deutsche Forschungsgemeinschaft (DFG) within the priority programme SpinCat (RE 1052/24-1) and the Bundesministerium f\"u{}r Bildung und Forschung (BMBF).

\section{References}

\bibliographystyle{apsrmp4-1}
\bibliography{bibfile}

\end{document}